\documentclass[aps, preprint, amsmath, amssymb, nofootinbib, 11pt]{revtex4-1}

\usepackage[utf8]{inputenc}
\usepackage[T1]{fontenc}
\usepackage{url, booktabs, amsfonts, nicefrac, microtype}
\usepackage{graphicx, color, slashed, textcomp, bbm, mathdots, multirow, array}
\usepackage{natbib, revsymb4-2, bm, mathrsfs}
\usepackage[colorlinks=true, linkcolor=red, urlcolor=blue, citecolor=blue]{hyperref}
\usepackage{cleveref, subcaption, makecell}
\usepackage[normalem]{ulem}
\usepackage{listings, xcolor}

\graphicspath{{figures/}}

\begin{document}

\title{Environmental Imprints of Dark Matter and Accretion Disks on Eccentric EMRIs around Kerr Black Holes}

\author{Hai-Chao Yuan$^{a}$}
\author{Yong Tang$^{a,b,c}$}
\affiliation{
$^{a}$School of Astronomy and Space Science, University of Chinese Academy of Sciences (UCAS), Beijing 100049, China\\
$^{b}$School of Fundamental Physics and Mathematical Sciences, Hangzhou Institute for Advanced Study, UCAS, Hangzhou 310024, China\\
$^{c}$International Center for Theoretical Physics Asia-Pacific, Beijing/Hangzhou, China
}

\date{\today}

\begin{abstract}
Extreme-mass-ratio inspirals (EMRIs), systems in which a stellar-mass compact object spirals into a supermassive black hole ($M_1$), are prime targets for the space-based gravitational-wave interferometers and are acutely sensitive to their astrophysical environment. We quantify the joint imprint of a dark matter (DM) spike and an accretion disk on the waveforms of eccentric, equatorial Kerr EMRIs, integrating four modular effects---DM dynamical friction, DM self-gravity, Newtonian accretion-disk torques, and disk self-gravity---into a fifth post-Newtonian (5PN) augmented analytic kludge (AAK) waveform model, and assess their measurability via a Fisher-matrix analysis. The DM dynamical-friction dephasing peaks at $M_1\sim 10^6\,M_\odot$, reaching $\sim 10^2$--$10^3\,\mathrm{rad}$; there the DM spike slope is measurable to sub-percent precision and both DM channels are unambiguously detectable, while omitting DM biases the intrinsic parameters significantly. At $M_1\gtrsim 10^8\,M_\odot$ the DM and disk self-gravity channels dominate their dissipative counterparts. The disk parameters remain degenerate across the explored parameter space, since the supersonic regime dominates the inspiral and the subsonic-to-supersonic transition only mildly weakens this degeneracy without lifting it; a combined DM$+$disk analysis further reveals a cross-sector degeneracy between the DM slope and the disk parameters. This physical degeneracy, reflected in the near-singular Fisher matrix, implies that independently measuring the disk parameters requires a relativistic torque model or an electromagnetic counterpart.
\end{abstract}

\maketitle

\section{Introduction}\label{sec:intro}

The advent of gravitational-wave (GW) astronomy~\cite{Abbott:2016gw} has opened a new observational window onto the Universe. The forthcoming LISA~\cite{AmaroSeoane:2017LISA}, Taiji~\cite{Hu:2017taiji}, and TianQin~\cite{Fan:2022tianqin} will extend this reach to the millihertz band, where EMRIs are among the most anticipated sources. An EMRI consists of a stellar-mass compact object of mass $m_2 \sim 1$--$10^2\,M_\odot$ slowly inspiraling into a supermassive black hole (SMBH) of mass $M_1 \sim 10^5$--$10^{10}\,M_\odot$. Owing to the extreme mass ratio $\epsilon \equiv m_2/M_1 \ll 1$, the inspiral proceeds over $\sim 10^4$--$10^5$ orbital cycles in the LISA sensitivity band, during which the secondary acts as a precision probe of the central black hole's spacetime geometry~\cite{Barack:2004ak, Babak:2017EMRI}. This prolonged interaction with the strong-field regime makes EMRIs ideal laboratories for testing general relativity, measuring black hole masses and spins with exquisite accuracy, and constraining the astrophysical environment of galactic nuclei~\cite{Gair:2004EMRI, AmaroSeoane:2007EMRI, Mancieri2026, Yang:2026QOS}.

A distinctive feature of EMRIs is their acute sensitivity to environmental perturbations. Any additional dissipative or conservative force, even if orders of magnitude weaker than the leading-order GW back-reaction, can accumulate over millions of radians of GW phase to produce a detectable imprint~\cite{Barack:2004ak}. This sensitivity has motivated extensive work on environmental effects in GW sources~\cite{Bertone:2020DMGW,Cardoso:2022envPRL,Cole:2023distinguish,Barausse:2014env,2025PhRvD.111h3010C,Chen:2026inner}. Galactic nuclei---the natural habitats of SMBHs---can harbor two physically distinct environments that may coexist: (i) a dense DM spike, formed through the adiabatic contraction of the pre-existing DM halo during the SMBH's growth~\cite{Gondolo:1999ef,Nishikawa:2017rjs,Zhang:2025dm}, and (ii) a geometrically thin, radiatively efficient accretion disk, fueled by the gas supply in active galactic nuclei (AGN)~\cite{Shakura:1973disks,Kocsis_2011}, though stellar interactions and recoil kicks can regulate the formation rate of disk-embedded EMRIs~\cite{Xue:2026recoil}. Characterizing the influence of these environments is a pressing task: on one hand, it offers a unique opportunity to probe DM microphysics and accretion-disk astrophysics through purely gravitational observations; on the other, neglecting such effects risks introducing systematic biases in parameter estimation~\cite{Speri:2023accretion, Duque:2025eccentric}.

The gravitational influence of an accretion disk on an EMRI has been studied extensively. Early work addressed eccentricity and inclination evolution in thin disks~\cite{Levin:2007emi} and the role of dynamical friction and migration in gas-rich environments~\cite{Derdzinski:2021disk, Zwick:2022disk, Zwick:2023disk, Garg:2022disk, SanchezSalcedo2020}. Ref.~\cite{Kocsis_2011} provided a comprehensive assessment, establishing that non-axisymmetric gravitational torques---migration---dominate over hydrodynamic drag and disk self-gravity. Ref.~\cite{Yunes:2011DM} showed that this migration imprints an observable dephasing on the EMRI waveform. Ref.~\cite{Speri:2023accretion} performed a Bayesian parameter-estimation study of migration torques for circular EMRIs using the FastEMRIWaveform (FEW) package~\cite{Katz:2021few,2021PhRvL.126e1102C}, demonstrating that LISA can measure the torque amplitude and that ignoring migration biases the primary mass and spin. Ref.~\cite{Duque:2025eccentric} extended this framework to eccentric inspirals on a Schwarzschild background, introducing unified phenomenological matching formulas that interpolate between subsonic type-I migration ($e \ll h$) and supersonic local dynamical friction ($e \gtrsim h$), where $e$ is the orbital eccentricity and $h$ is the disk aspect ratio, and showing that the transonic transition enables independent measurements of disk viscosity and accretion rate. Ref.~\cite{Fantoccoli:2026disk} performed a fully Bayesian analysis of eccentric EMRIs on a Kerr background using relativistic disk-torque models, finding that the relativistic corrections break the disk-parameter degeneracy inherent to Newtonian torque models and thereby enable independent measurement of the disk parameters without an electromagnetic counterpart. EMRIs and IMRIs embedded in AGN accretion disks have been studied~\cite{Peng:2023sync,Peng:2025fate}; disk-induced gravitational-wave phase shifts~\cite{Tagawa:2026shift} and EMRI measurements of the disk environment~\cite{Liu:2026AGN} have been investigated. Relatedly, Ref.~\cite{Zhang:2026diskobs} studied the images of thin accretion disks around rotating black holes embedded in DM halos, leaving their GW imprint unexplored.

The impact of DM spikes on EMRI dynamics has been studied through a parallel line of investigation. The density enhancement from adiabatic contraction~\cite{Gondolo:1999ef}---subsequently refined with general-relativistic corrections~\cite{Sadeghian:2013spike,Ferrer:2017dm}---can reach $\rho_{\rm DM} \sim 10^6$--$10^{18}\,\mathrm{GeV\,cm^{-3}}$ near the SMBH~\cite{Zhang:2025dm,Shen:2024GCspike,Hu:2023pulsar}. Dynamical friction from the DM background constitutes a potentially significant dissipative channel~\cite{Eda:2013DM,Macedo:2013DM,Yue:2018minispike,Yue:2019catalyst}, with recent work incorporating the velocity distribution of DM particles and its effect on orbital eccentricity and waveform dephasing~\cite{Becker:2022DM,Li:2022DMspike,Chen:2025SIDM}. The enclosed DM mass also modifies the local gravitational potential, enhancing the effective central mass that governs the radiation-reaction timescale~\cite{Zhang:2025dm}. Ref.~\cite{Wu:2026DMdisk} studied the phase distinguishability between DM and accretion-disk environments for eccentric EMRIs on a Schwarzschild background, finding that eccentricity serves as an auxiliary diagnostic in addition to the observation duration. The effect of dark-matter spikes on gravitational-wave signals has been studied across a range of systems~\cite{Shen:2025SGWB,Hu:2025GWB,Li:2025DM,Xu:2026LMCM31}. Environmental perturbations beyond a spherical spike have also been treated in a unified framework~\cite{Jiang:2023dirty,Jiang:2024dirty}. To date, however, no implementation has combined these DM effects with accretion-disk torques into a single waveform-generation framework on a Kerr background, despite the fact that both environments may coexist in the same galactic nucleus~\cite{Becker:2023comparison}. Furthermore, existing DM studies have largely been restricted to Schwarzschild backgrounds or circular orbits, so that the interplay between Kerr spacetime geometry and environmental perturbations remains unexplored.

In this paper, we investigate eccentric, equatorial EMRIs around Kerr black holes embedded in both a DM spike and an accretion disk---a prograde configuration characteristic of disk-embedded EMRIs. Unlike existing studies that have treated DM effects in Schwarzschild spacetimes~\cite{Eda:2013DM,Macedo:2013DM,Yue:2018minispike,Becker:2022DM,Li:2022DMspike} or disk torques on Schwarzschild eccentric backgrounds~\cite{Duque:2025eccentric}, we present a combined analysis of both environmental sectors on a Kerr 5PN AAK inspiral baseline, complementing recent studies that have separately addressed DM~\cite{Wu:2026DMdisk} and accretion-disk~\cite{Fantoccoli:2026disk} effects on eccentric EMRIs. To enable systematic parameter-space surveys, we integrate four environmental effects as modular extensions within the FEW package~\cite{Katz:2021few,2021PhRvL.126e1102C}: DM dynamical friction, incorporating the velocity-dependent corrections $N_1$ and $N_2$ to the Chandrasekhar formula (Sec.~\ref{sec:DM_DF}); DM self-gravity, treated as a leading-order effective-mass correction to the radiation-reaction fluxes, the orbital frequencies, and the flux-to-elements Jacobian (Sec.~\ref{sec:selfgrav}); accretion-disk torques, modeled via the unified phenomenological matching formulas of Ref.~\cite{Duque:2025eccentric} (Sec.~\ref{sec:disk}); and disk self-gravity, included via the same effective-mass approach. All perturbations are incorporated as orbit-averaged additions to the vacuum 5PN Kerr inspiral, with each module independently switchable. We apply this framework to a fiducial $M_1=10^6\,M_\odot$ EMRI, mapping the dephasing across the parameter space of primary mass, initial separation, eccentricity, spin, and DM spike slope, and performing a Fisher-matrix analysis~\cite{Cutler:1994ys,Cutler:2007mi} to assess the measurability of the environmental parameters and the biases incurred when they are neglected. Our analysis reveals that Newtonian disk torques in a Kerr background effectively provide only one environmental degree of freedom---the disk surface density and aspect ratio remain degenerate across the explored parameter space, which finds that this degeneracy is only lifted by relativistic torque corrections.

The paper is organized as follows. Section~\ref{sec:formulation} presents the physical models for the DM spike, the dynamical-friction force, the self-gravity prescriptions, and the accretion-disk torque model. Section~\ref{sec:analysis} describes the integration of these effects into the FEW pipeline and the Fisher-matrix methodology. Results for a fiducial $M_1=10^6\,M_\odot$ system are presented in Sec.~\ref{sec:results}, and we summarize our conclusions in Sec.~\ref{sec:conclusions}.

\section{Formulation}\label{sec:formulation}

\subsection{Dark Matter Dynamical Friction}\label{sec:DM_DF}

We consider a scenario in which the central black hole grew adiabatically inside a pre-existing Navarro--Frenk--White (NFW) halo, giving rise to a relativistic DM spike~\cite{Gondolo:1999ef,Nishikawa:2017rjs,Zhang:2025dm}. The density profile, generalized to a Kerr black hole by replacing the Schwarzschild innermost stable circular orbit (ISCO) with the equatorial innermost bound orbit, is
\begin{equation}\label{eq:rho_spike}
\rho_{\rm DM}(r) = \rho_{\rm sp} \left(1 - \frac{r_{\rm mb}}{r}\right)^3 \left(\frac{r}{r_{\rm sp}}\right)^{-\xi},\qquad r \ge r_{\rm mb},
\end{equation}
and $\rho_{\rm DM}=0$ for $r< r_{\rm mb}$, where
\begin{equation}
r_{\rm mb} = 2 - a + 2\sqrt{1-a}
\end{equation}
is the Kerr equatorial innermost bound orbit radius in units of $R_g\equiv GM_1/c^2$~\cite{Misner1973}, and $a$ is the dimensionless black-hole spin. For a Schwarzschild black hole ($a=0$), $r_{\rm mb}=4$ in units of $GM_1/c^2$, which equals $2R_s$ with $R_s=2GM_1/c^2$, recovering the factor $(1-2R_s/r)^3$. The power-law index $\xi$ is related to the initial NFW inner slope $\gamma$ ($0<\gamma<2$), hereafter the DM spike slope, by
\begin{equation}\label{eq:xi}
\xi = \frac{9-2\gamma}{4-\gamma},
\end{equation}
yielding $\xi=2.25,\,7/3,\,2.5$ for $\gamma=0,1,2$, respectively.

\begin{figure}
    \centering
    \includegraphics[width=\textwidth]{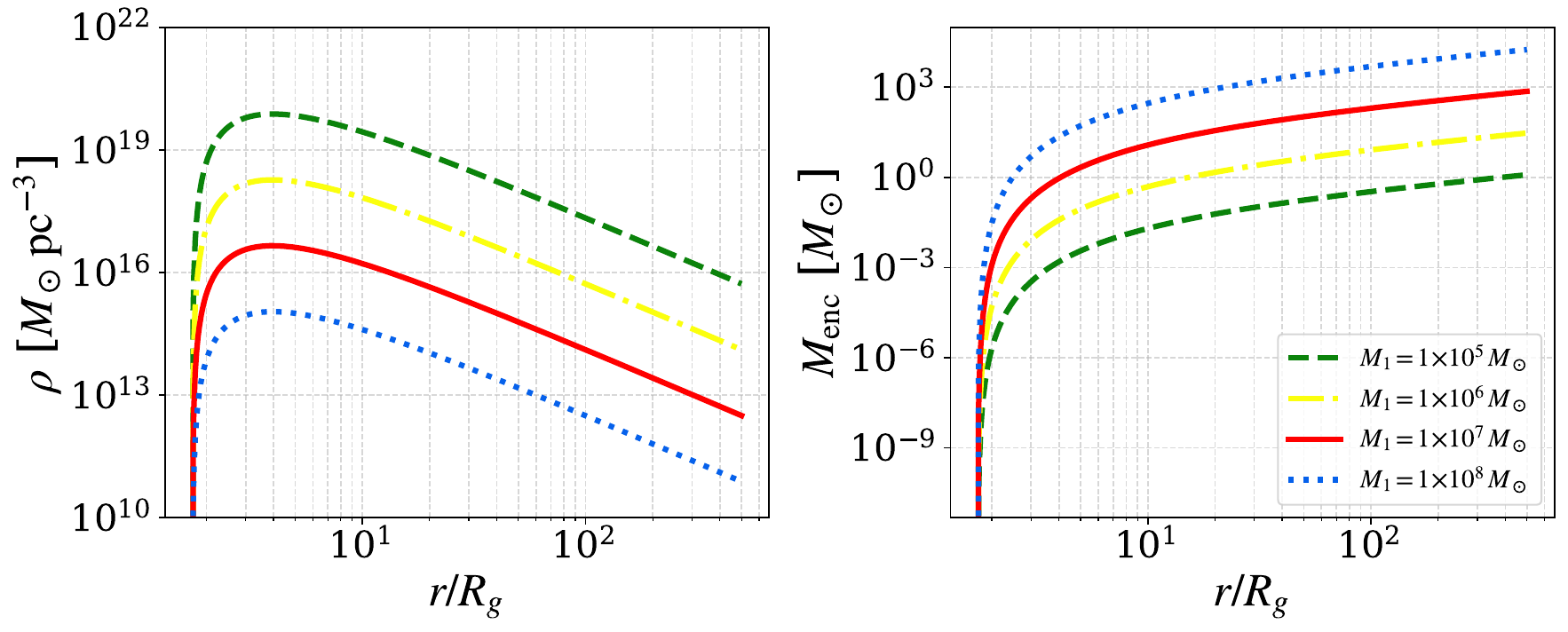}
    \caption[DM spike density and enclosed mass for different primary masses]{
        Density (left) and enclosed mass $M_{\rm enc}(r)$ (right) of the DM spike for four primary masses spanning the EMRI regime ($M_1 = 10^5$--$10^8\,M_\odot$, spin $a=0.9$, and $\gamma=1$). Radii are in units of $R_g = GM_1/c^2$.
    }
    \label{fig:dm_spike_mass}
\end{figure}

The normalization $\rho_{\rm sp}$ and scale radius $r_{\rm sp}$ are obtained from the initial NFW parameters $(\rho_0,r_0)$ and the black hole mass $M_1$ via~\cite{Gondolo:1999ef}
\begin{align}
\rho_{\rm sp} &= \rho_0 \left(\frac{r_{\rm sp}}{r_0}\right)^{-\gamma},\label{eq:rhosp}\\
r_{\rm sp} &= \alpha_\gamma\, r_0 \left(\frac{M_1}{\rho_0 r_0^3}\right)^{1/(3-\gamma)},\label{eq:rsp}
\end{align}
where $\alpha_\gamma$ is an adiabatic contraction constant (e.g., $\alpha_{\gamma=1}=0.122$). The NFW parameters $(\rho_0,r_0)$ are not free but are fixed by the $M_1$--$\sigma_\star$ relation, where $\sigma_\star$ is the stellar velocity dispersion of the host galaxy, and the virial definition of the halo, together with the mass--concentration relation~\cite{Gultekin:2009fg, SanchezConde:2014pwa, Zhang:2025dm}. For $M_1=10^6\,M_\odot$ ($\gamma=1$) we calculate $\rho_0\simeq4.0\times10^{-25}\,\mathrm{g\,cm^{-3}}$ and $r_0\simeq7.6\times10^3\,\mathrm{pc}$. The resulting anti-correlation between $M_1$ and $\rho_{\rm DM}$ in Fig.~\ref{fig:dm_spike_mass}, as we shall show later (Sec.~\ref{sec:results}), is the dominant source of the mass dependence of the DM dephasing.

The DM density $\rho_{\rm DM}(r)$ of Eq.~\eqref{eq:rho_spike} contributes to the dynamical-friction force. The gravitational wake of the secondary moving through the DM spike leads to a drag force $\mathbf{F}_{\rm DF}$, where the subscript DF denotes dynamical friction. For an isotropic Maxwellian velocity distribution, the Chandrasekhar formula~\cite{Chandrasekhar:1943df, Zhang:2025dm} generalized to include high-speed corrections reads
\begin{equation}\label{eq:DF}
\mathbf{F}_{\rm DF} = -\,\frac{4\pi G^2 m_2^2\,\rho_{\rm DM}(r)}{v^2}\,
\bigl( \ln\Lambda\,N_1 + N_2 \bigr)\,
\frac{\mathbf{v}}{v},
\end{equation}
where $v=|\mathbf{v}|$ and $\ln\Lambda\simeq10$ is the Coulomb logarithm. The functions $N_1$ and $N_2$ describe, respectively, the fraction of DM particles moving slower than $v$ and the contribution of the faster-moving particles. Following the standard isotropic treatment~\cite{Binney:2008galactic, Zhang:2025dm}, we take a velocity dispersion $\sigma_v = 0.5\,v_{\rm esc}$, as expected for an isotropic velocity distribution in the near-isothermal spike profile where the one-dimensional dispersion satisfies $\sigma_v\approx v_{\rm esc}/2$ in a Keplerian potential, with $v_{\rm esc}=\sqrt{2GM_1/r}$; these functions are given by
\begin{align}
N_1 &= \int_0^{v} f(v')\,dv',\\
N_2 &= \int_{v}^{v_{\rm esc}} f(v')\left[\ln\!\left(\frac{v'+v}{v'-v}\right) - 2\frac{v}{v'}\right] dv',\label{eq:N2}
\end{align}
with $f(v) = 4\pi v^2 (2\pi\sigma_v^2)^{-3/2} \exp(-v^2/2\sigma_v^2)$. Eq.~\eqref{eq:DF} is the Chandrasekhar dynamical-friction force including the high-speed correction $N_2$, which accounts for DM particles moving faster than the secondary.

To incorporate the dynamical-friction force into the long-term orbital evolution we use the Gaussian perturbation equations~\cite{PoissonWill:2014}. For a perturbing force with radial and tangential components $R$ and $S$ (positive outward and forward, respectively),
\begin{align}
\frac{dp}{df} &= \frac{2p^3}{GM_1 (1+e\cos f)^3}\,\frac{S}{m_2},\label{eq:gauss_p}\\
\frac{de}{df} &= \frac{p^2}{GM_1}\left[ \frac{\sin f}{(1+e\cos f)^2}\,\frac{R}{m_2} + \frac{2\cos f + e(1+\cos^2 f)}{(1+e\cos f)^3}\,\frac{S}{m_2}\right].\label{eq:gauss_e}
\end{align}
Here $p$, $e$, and $f$ are the semi-latus rectum, eccentricity, and true anomaly, respectively, and $R = -F_{\rm DF}\,v_r/v$ and $S = -F_{\rm DF}\,v_\phi/v$, where $v_r$ and $v_\phi$ are the radial and azimuthal velocity components and $F_{\rm DF}=|\mathbf{F}_{\rm DF}|$ is the magnitude of the dynamical-friction force. The division by $m_2$ converts the force into an acceleration.

The instantaneous rates are orbit-averaged over one orbital period, and the resulting orbit-averaged rates $\dot p$ and $\dot e$ are added to the vacuum 5PN rates provided by FEW.

\subsection{Dark Matter Self-Gravity}\label{sec:selfgrav}

The DM spike not only exerts dynamical friction on the secondary, but also enhances the local gravitational field through its enclosed mass. For a static, spherically symmetric DM distribution, the DM particles do not contribute directly to the GW emission---their intrinsic quadrupole moment vanishes identically. However, the enclosed mass modifies the Keplerian relation between the orbital frequency and separation. In the Newtonian limit, a test particle at radius $r$ orbits with angular frequency
\begin{equation}
\omega^2 = \frac{G M_{\rm eff}(r)}{r^3},\qquad
M_{\rm eff}(r) = M_1 + M_{\rm enc}(r),
\end{equation}
where $M_{\rm enc}(r)$ is the enclosed mass within radius $r$. For eccentric orbits, the instantaneous enclosed mass varies along the orbit; we therefore compute the orbit-averaged value $\langle M_{\rm enc}\rangle$ using the same orbit-averaging procedure, which gives a representative effective mass for the radiation-reaction calculation. The integration extends from the Kerr innermost bound orbit radius $r_{\rm mb}=2-a+2\sqrt{1-a}$ (in units of $R_g$) out to the instantaneous orbital radius.

Since the quadrupole formula $\mathcal{L}_{\rm GW} \propto \mu^2 r^4 \omega^6$, with $\mu=M_1m_2/(M_1+m_2)\approx m_2$ the reduced mass, retains its functional form (the DM is static and spherically symmetric, contributing no additional time-varying multipoles), the only modification enters through the replacement $M_1 \to M_{\rm eff}$ when eliminating $r$ via Kepler's law. At leading (0PN) order, the orbit-averaged energy and angular momentum fluxes become
\begin{align}
\langle \dot E \rangle_{\rm DM} &= -\frac{32}{5}\frac{G^4 m_2^2 M_{\rm eff}^3}{c^5 a_*^5} F(e),\label{eq:Edot_DM}\\
\langle \dot L_z \rangle_{\rm DM} &= -\frac{32}{5}\frac{G^{7/2} m_2^2 M_{\rm eff}^{5/2}}{c^5 a_*^{7/2}} G(e),\label{eq:Lzdot_DM}
\end{align}
where $a_* = p/(1-e^2)$ is the semi-major axis and $F(e)$ and $G(e)$ are the standard Peters--Mathews eccentricity enhancement functions~\cite{Peters:1963ux,Peters:1964zz}. The corresponding correction factors relative to the vacuum values are
\begin{equation}
\mathcal{R}_E \equiv \frac{\langle\dot E\rangle_{\rm DM}}{\langle\dot E\rangle_{\rm 0PN}}
= \left(\frac{M_{\rm eff}}{M_1}\right)^3,\qquad
\mathcal{R}_L \equiv \frac{\langle\dot L_z\rangle_{\rm DM}}{\langle\dot L_z\rangle_{\rm 0PN}}
= \left(\frac{M_{\rm eff}}{M_1}\right)^{5/2}.
\end{equation}
For $\delta m \equiv M_{\rm eff} - M_1 \simeq \langle M_{\rm enc}\rangle \ll M_1$, these reduce to $\mathcal{R}_E \simeq 1 + 3\,\delta m/M_1$ and $\mathcal{R}_L \simeq 1 + (5/2)\,\delta m/M_1$.

The flux rescaling above is the dissipative imprint of the enclosed mass. Two conservative channels of the same order accompany it. First, at fixed orbital elements the fundamental frequencies that drive the wave phases are shifted: at leading order all frequencies acquire the common factor $\sqrt{M_{\rm eff}/M_1}$, while the radial frequency receives an additional gradient contribution,
\begin{equation}\label{eq:SG_freq}
\Omega_{\phi,\theta} \to \Omega_{\phi,\theta}\,\sqrt{\frac{M_{\rm eff}}{M_1}},\qquad
\Omega_{r} \to \Omega_{r}\,\sqrt{\frac{M_{\rm eff}}{M_1}} \;+\; \Omega_{\phi}\,\sqrt{\frac{M_{\rm eff}}{M_1}}\left(1 - \sqrt{1+\frac{\langle 4\pi r^3 \rho_{\rm DM}\rangle}{M_{\rm eff}}}\right),
\end{equation}
where the second term represents the classical apsidal regression of an extended mass, derived from $\Omega_r = \Omega_\phi - \kappa$ with $\kappa^2 = \Omega^2(1 + d\ln M_{\rm eff}/d\ln r)$; the gradient correction enters only the Newtonian epicyclic part and thus appears as an additive shift rather than a multiplicative factor on $\Omega_r$. The common frequency scaling advances the accumulated phase over the inspiral, while the gradient term reduces the periastron advance---the classical apsidal regression---and is of the same order as the dissipative channel.

Second, the conversion of the fluxes into rates of the orbital elements involves the vacuum Kerr relations $E(p,e)$ and $L_z(p,e)$, which are themselves modified by the enclosed mass. Expanding the effective Jacobian to first order in $\delta m$, the correction to the element rates takes the same difference form as the flux replacement,
\begin{equation}\label{eq:SG_jac}
\delta\!\begin{pmatrix}\dot p \\ \dot e\end{pmatrix}
= \left[\mathbf{J}_{\rm N}^{-1}(M_{\rm eff}) - \mathbf{J}_{\rm N}^{-1}(M_1)\right]
\begin{pmatrix}\dot E \\ \dot L_z\end{pmatrix}_{\rm 0PN},
\end{equation}
where $\mathbf{J}_{\rm N}$ is the Jacobian of the Newtonian relations $E_{\rm N} = -G M_{\rm eff}\,m_2(1-e^2)/(2p)$ and $L_{z,{\rm N}} = m_2\sqrt{G M_{\rm eff}\,p}$, evaluated including the gradients $\partial\langle M_{\rm enc}\rangle/\partial p$ and $\partial\langle M_{\rm enc}\rangle/\partial e$ of the orbit-averaged enclosed mass, and the correction is added to the output of the vacuum Kerr transformation.

In our numerical implementation, the 5PN adiabatic fluxes provided by FEW are decomposed into their 0PN baseline and higher-order PN corrections: only the 0PN part is replaced by the DM-corrected version~\eqref{eq:Edot_DM}--\eqref{eq:Lzdot_DM}, the fundamental frequencies are rescaled according to Eq.~\eqref{eq:SG_freq}, and the Jacobian correction of Eq.~\eqref{eq:SG_jac} is applied on top of the vacuum Kerr transformation. All three leading-order appearances of $M_{\rm eff}$ are thus treated consistently; the residual inconsistency stems only from the uncorrected higher-order PN terms and is suppressed by an additional factor of $v^2/c^2 \sim 0.01$--$0.1$ relative to the leading DM correction, i.e.\ $\mathcal{O}[(\delta m/M_1)\,v^2/c^2]$---negligible at current sensitivity levels.

\subsection{Accretion Disk Torques}\label{sec:disk}

A non-negligible fraction of EMRIs are expected to reside in the accretion disks of AGN~\cite{Speri:2023accretion,Duque:2025eccentric,Zeng:2026captured}. Such disks exchange energy and angular momentum with the secondary through hydrodynamical and gravitational interactions, altering the inspiral trajectory~\cite{Kocsis_2011}. We adopt a radiatively efficient, geometrically thin, stationary $\alpha$-disk model~\cite{Shakura:1973disks}, parameterized by the surface density and aspect ratio power laws, with $H$ the disk scale height:
\begin{equation}
\Sigma(r) = \Sigma_0 \left(\frac{r}{10\,GM_1/c^2}\right)^{-\Sigma_p},\qquad
h(r) \equiv \frac{H}{r} = h_0 \left(\frac{r}{10\,GM_1/c^2}\right)^{(2\Sigma_p-1)/4},
\end{equation}
where the exponent of $h(r)$ ensures a steady-state mass accretion rate. For the $\alpha$-disk inner region ($r \lesssim 10^2\,GM_1/c^2$), radiation pressure dominates with $\Sigma_p = -3/2$, giving $\Sigma \propto r^{3/2}$ and $h \propto r^{-1}$. The parameters $\Sigma_0$ and $h_0$ encapsulate the disk microphysics and can be mapped to the Shakura--Sunyaev viscosity coefficient $\alpha$ and the Eddington-scaled accretion rate $f_{\rm Edd}$ through Eqs.\ (4)--(5) of Ref.~\cite{Duque:2025eccentric}. The disk is considered equatorial and aligned with the primary's spin axis~\cite{Bardeen:1975aligned}, and we restrict to prograde orbits.

The nature of the disk--secondary interaction is governed by the Mach number of the relative motion, $\Delta v/c_s$, where $c_s = h\,v_{\rm K}$ is the isothermal sound speed, $v_{\rm K}=r\Omega_{\rm K}$ the Keplerian velocity, and $\Delta v \simeq e\,v_{\rm K}$ is the relative velocity between the secondary and the local gas, which for an eccentric orbit is set by the eccentric radial motion, so that $\Delta v/c_s \simeq e/h$~\cite{Duque:2025eccentric}. Two distinct regimes arise. When $e \ll h$, the secondary moves subsonically and excites spiral density waves in the disk; these waves resonate at Lindblad radii and back-react on the secondary with a net gravitational torque---a process known as type-I planetary migration~\cite{Tanaka:2002migration,Tanaka:2004migration,Cresswell:2008migration,FairbairnRafikov2025}. When $e \gg h$, the motion becomes supersonic and the spiral density wave structure can no longer form; the dominant interaction switches to local dynamical friction, where the gravitational wake trailing the secondary exerts a drag force~\cite{Ostriker:1999df}. Both effects operate on a characteristic inverse timescale
\begin{equation}
\frac{1}{t_{\rm gas}} = \epsilon\,\frac{\Sigma\,a_*^2\,\Omega_{\rm K}}{M_1\,h^4},
\end{equation}
where $\Omega_{\rm K}=\sqrt{GM_1/a_*^3}$.

The transonic region $e \sim h$, where the linear approximations underlying both analytic limits break down due to shock formation, is bridged by the phenomenological matching formulas of Ref.~\cite{Duque:2025eccentric}, which give the inverse timescales $1/t_e$ for eccentricity damping and $1/t_a$ for semi-major axis migration:
\begin{align}
\frac{1}{t_e} &= \frac{1}{t_{\rm gas}}\,\frac{0.78\,(1-e^2)^{1/4}}{1 + \dfrac{1}{30}\!\left(\dfrac{e}{h}\right)^3},\label{eq:te_match}\\
\frac{1}{t_a} &= 2\,C_{\rm sub}\,h^2\,\frac{1}{t_{\rm gas}}\,(1-e^2)\,
\frac{1 - \left(\dfrac{e}{1.25\,h}\right)^4}{1 + \left(\dfrac{e}{1.75\,h}\right)^5},\label{eq:ta_match}
\end{align}
with $C_{\rm sub}=2.15+0.04\,\Sigma_p$ the subsonic migration torque coefficient~\cite{Tanaka:2024okada}. These expressions smoothly interpolate between the subsonic and supersonic regimes, providing a unified torque prescription valid for arbitrary eccentricities. The physical-time rates are $\dot a_* = -a_*/t_a$ and $\dot e = -e/t_e$, converted to the semi-latus rectum via $\dot p = (1-e^2)\dot a_* - 2a_* e \dot e$ and then to FEW's scaled time. The formulas reduce to the type-I migration rate in the subsonic limit $e\ll h$ and to the supersonic dynamical-friction rate $\propto (h/e)^3$ in the limit $e\gg h$, as expected from the construction of Ref.~\cite{Duque:2025eccentric}.

Disk self-gravity is included as an optional feature via the same effective-mass approach as for the DM spike (Sec.~\ref{sec:selfgrav}), with the orbit-averaged enclosed mass $\langle M_{\rm enc}\rangle$ computed from the equatorial surface density $\Sigma$ using the same orbit-averaging procedure, the radial integral extending from the disk's inner edge at the separatrix pericenter $p_{\rm sep}/(1+e)$ out to the instantaneous orbital radius, where $p_{\rm sep}(a, e, x_I)$ is the Kerr separatrix (the boundary between bound and plunging orbits), with $x_I\equiv\cos I$ the orbital-inclination cosine.

\section{Numerical Implementation and Fisher Matrix}\label{sec:analysis}

The environmental effects formulated in Sec.~\ref{sec:formulation} are incorporated into the FEW framework~\cite{Katz:2021few,2021PhRvL.126e1102C,2025PhRvD.112j4023C} as orbit-averaged additions to the adiabatic inspiral trajectory. FEW computes the vacuum inspiral rates in energy and angular momentum, $(\dot E_{\rm 5PN}, \dot L_{z\,\rm 5PN})$, at 5PN order for generic Kerr orbits, using analytic flux formulas~\cite{Fujita:2020pn}. These fluxes are transformed to the time derivatives of the quasi-Keplerian orbital elements $(\dot p, \dot e, \dot Y)$ via a Jacobian that respects the Kerr geodesic relations between the orbital constants $(E, L_z)$ and $(p, e, x_I)$. Here $Y\equiv\cos\iota=L/\sqrt{L^2+Q}$ is the post-Newtonian inclination cosine, where $L$ and $Q$ are the specific orbital angular momentum and the Carter constant; for the equatorial orbits considered here, $x_I=Y=1$.

The environmental corrections are computed independently: on the DM side through orbit-averaged Gaussian perturbation equations for dynamical friction (Sec.~\ref{sec:DM_DF}) and 0PN flux rescaling factors $(M_{\rm eff}/M_1)^3$ and $(M_{\rm eff}/M_1)^{5/2}$ together with the leading-order conservative corrections to the fundamental frequencies and to the flux-to-elements Jacobian for self-gravity (Sec.~\ref{sec:selfgrav}); on the accretion-disk side through the phenomenological matched torque timescales $t_a$ and $t_e$, with disk self-gravity included via the same flux-rescaling approach (Sec.~\ref{sec:disk}). The total rates are obtained by superposition in FEW's scaled time $\tau$,
\begin{align}
\dot p(\tau) &= \dot p_{\rm 5PN}(\tau) + \dot p_{\rm DM}(\tau) + \dot p_{\rm disk}(\tau),\label{eq:pdot_total}\\
\dot e(\tau) &= \dot e_{\rm 5PN}(\tau) + \dot e_{\rm DM}(\tau) + \dot e_{\rm disk}(\tau),\label{eq:edot_total}
\end{align}
with separate flags controlling each environmental contribution. The trajectory is integrated using an adaptive eighth-order Runge--Kutta solver until $p$ reaches the Kerr separatrix $p_{\rm sep}(a, e, x_I)$.

The environmental perturbations are treated as secular corrections on the radiation-reaction timescale, justified by the two-timescale nature of EMRIs: the orbital phases $\Phi_{\phi,\theta,r}$ vary on the fast orbital period, while the orbital elements $(p, e, Y)$ evolve on the much slower inspiral timescale.
The numerical validation of this orbit-averaging framework is presented in Sec.~\ref{sec:results} (Fig.~\ref{fig:adiabatic_disk}).

All waveform computations in this work use the LISA detector response, which applies the time-dependent antenna patterns $F_{+,\times}$ and the associated Doppler modulation to produce the detector output $h_{\rm I}(t) = (\sqrt{3}/2)(F_+ h_+ + F_\times h_\times)$, where the subscript I denotes the interferometric channel. The dephasing $\Delta\phi_{22}(t)$, defined in Eq.~\eqref{eq:dephasing}, is extracted from the inspiral trajectory via $\Phi_\phi(t)$ and is therefore independent of the detector response. For the Fisher-matrix analysis and mismatch calculations, the noise-weighted inner products $\langle\cdot|\cdot\rangle$ in Eqs.~\eqref{eq:mismatch}--\eqref{eq:Gamma} are evaluated on the detector-frame strain $h_{\rm I}(t)$, which incorporates the correct LISA sensitivity. For generic Kerr inspirals we employ the AAK waveform model~\cite{Babak:2007nk, Chua:2015aak, Chua:2017aak}, which maps the evolving quasi-Keplerian elements to GW harmonics via the Kerr fundamental frequencies $(\Omega_\phi, \Omega_\theta, \Omega_r)$ and a leading-order spin-weight $-2$ spheroidal harmonic decomposition. The AAK model computes the waveform amplitudes to leading (Newtonian quadrupole) order, while its phase evolution is driven by the 5PN Kerr frequency trajectories, ensuring that the accumulated dephasing from environmental effects is captured accurately. Following Ref.~\cite{Kocsis_2011}, we define the dephasing on the dominant $(l,m)=(2,2)$ harmonic. In the AAK waveform model the phase of this harmonic is $\Phi_{22}(t) = 2\Phi_\phi(t)$, where $\Phi_\phi(t)$ is the azimuthal orbital phase obtained directly from the inspiral trajectory. The environmental dephasing is then
\begin{equation}\label{eq:dephasing}
\Delta\phi_{22}(t) = 2\bigl[\Phi_\phi^{\rm env}(t) - \Phi_\phi^{\rm vac}(t)\bigr],
\end{equation}
evaluated at the end of the observation $t=T_{\rm obs}=4\,\mathrm{yr}$. For assessing the detectability of environmental effects we complement this geometric phase shift with the noise-weighted waveform mismatch $\mathcal{M}$, defined in Eq.~\eqref{eq:mismatch} below, which directly quantifies the statistical distinguishability of the environmental and vacuum signals.

The mismatch between two waveforms $h_1$ and $h_2$ is defined in terms of the noise-weighted inner product~Eq.~\eqref{eq:Gamma} as~\cite{Lindblom:2008}
\begin{equation}\label{eq:mismatch}
\mathcal{M}(h_1,h_2) = 1 - \max_{t_c}\,
\frac{|\langle h_1 | h_2\rangle|^2}{\langle h_1|h_1\rangle\,\langle h_2|h_2\rangle},
\end{equation}
where the maximization over the time shift $t_c$, with the phase $\phi_c$ absorbed by the absolute value, accounts for the fact that these extrinsic parameters are optimized in any realistic search. We compute the mismatch between the vacuum waveform and each environmental waveform, both generated in the detector frame. A mismatch $\mathcal{M} \gtrsim (\mathrm{SNR})^{-2}$, where $\mathrm{SNR}$ is the signal-to-noise ratio, indicates that the two signals are statistically distinguishable.

To assess the statistical measurability of the environmental parameters and the systematic biases incurred when such effects are neglected, we employ the Fisher-matrix formalism~\cite{Cutler:1994ys}. Under the assumptions of stationary, Gaussian detector noise and a sufficiently high SNR, the likelihood is well approximated by a multivariate Gaussian, with the Fisher matrix given by
\begin{equation}
\Gamma_{ij} = 4\,\operatorname{Re}\int_0^\infty \frac{1}{S_n(f)}\left(\frac{\partial \tilde h}{\partial\theta_i}\right)^* \left(\frac{\partial \tilde h}{\partial\theta_j}\right) df,
\label{eq:Gamma}
\end{equation}
where $\operatorname{Re}$ denotes the real part, $\tilde h(f)$ is the Fourier-domain waveform~\cite{Speri:2023fastfourier} and $S_n(f)$ is the one-sided LISA noise power spectral density (PSD)~\cite{Robson:2019}, including the galactic confusion noise for a four-year observation. The inverse yields the covariance matrix $\mathbf C = \Gamma^{-1}$, whose elements we denote $C_{ij}$; the $1\sigma$ statistical uncertainty on parameter $\theta_i$ is $\sigma_{\theta_i} = \sqrt{C_{ii}}$, with correlation coefficient $\mathcal C_{ij}=C_{ij}/\sqrt{C_{ii}C_{jj}}$. The condition number $\kappa(\Gamma)\equiv\|\Gamma\|\cdot\|\Gamma^{-1}\|$ (spectral norm) quantifies the stability of the matrix inversion, with $\kappa\gg1$ indicating near-singularity. Because $\kappa$ is a global measure, the reliability of each $1\sigma$ uncertainty is instead set by the eigenvector structure of $\Gamma$: parameters aligned with the flat direction (the smallest-eigenvalue eigenvector) inherit the near singularity, while the correlation coefficients $\mathcal C_{ij}$, which are determined by this eigenvector structure, remain meaningful whenever the finite-difference derivatives lie above the numerical noise floor. A near-perfect correlation ($|\mathcal C_{ij}|\approx1$) together with a large condition number---and, in the disk configuration, an analytical degeneracy of the waveform derivatives---then diagnoses a fundamental parameter degeneracy.

When an environmental signal is recovered with a vacuum template, the inferred parameters are systematically biased. Under the linear-signal approximation, this offset is estimated via the standard linear-signal bias formalism~\cite{Cutler:2007mi}:
\begin{equation}
\Delta\theta^i \approx \bigl(\Gamma_{\rm vac}^{-1}\bigr)^{ij} \left\langle \frac{\partial h_{\rm vac}}{\partial\theta_j} \Big| h_{\rm env} - h_{\rm vac} \right\rangle,
\label{eq:bias}
\end{equation}
where $\Gamma_{\rm vac}$ and $h_{\rm vac}$ are the Fisher matrix and waveform under the vacuum hypothesis, and $h_{\rm env}$ is the injected environmental signal. In the bias analysis, the vacuum hypothesis is generated by switching off the environmental modules entirely, so that the waveform difference entering Eq.~\eqref{eq:bias} is evaluated exactly. The standard linear-signal bias formalism assumes linearity near the best-fit point; when the recovered bias exceeds a few $\sigma$, the numerical magnitudes serve primarily to diagnose the inadequacy of the vacuum model. The SNR is defined as $\mathrm{SNR}^2 = \langle h_{\rm env}|h_{\rm env}\rangle = 4\int_0^\infty |\tilde h_{\rm env}(f)|^2/S_n(f)\,df$. The finite-difference step sizes of all parameters are validated individually by requiring that the overlap between derivatives evaluated at step sizes $\delta$ and $\delta/2$ exceeds $0.99$, and that the corresponding Fisher-diagonal uncertainties satisfy $|2\sigma(\delta)/\sigma(\delta/2)-1|<5\%$. The parameter set is restricted to the intrinsic EMRI parameters and the environmental degrees of freedom listed above; extrinsic parameters are held fixed at fiducial values, and all Fisher uncertainties reported below are rescaled to a fixed detector-frame SNR of 50 (obtained by adjusting the luminosity distance per configuration) to facilitate comparison.

\section{Numerical Results}\label{sec:results}

We consider equatorial, eccentric Kerr EMRIs with secondary mass $m_2=10\,M_\odot$ and total observation time $T_{\rm obs}=4\,\mathrm{yr}$, adopting the LISA noise PSD of Ref.~\cite{Robson:2019}. The environmental impact is quantified through both the single-harmonic dephasing $\Delta\phi_{22}$ and the noise-weighted mismatch $\mathcal{M}$, defined in Sec.~\ref{sec:analysis}. We first scan the primary mass $M_1$ over $10^5$--$5\times10^8\,M_\odot$ at several values of the initial semi-latus rectum $p_0$, then fix $M_1=10^6\,M_\odot$ to examine the dependence on $a$, the initial eccentricity $e_0$, and $\gamma$, and finally perform a Fisher-matrix analysis for the DM-only, disk-only, and combined configurations.

\begin{figure}
    \centering
    \includegraphics[width=1\textwidth, height=0.4\textwidth]{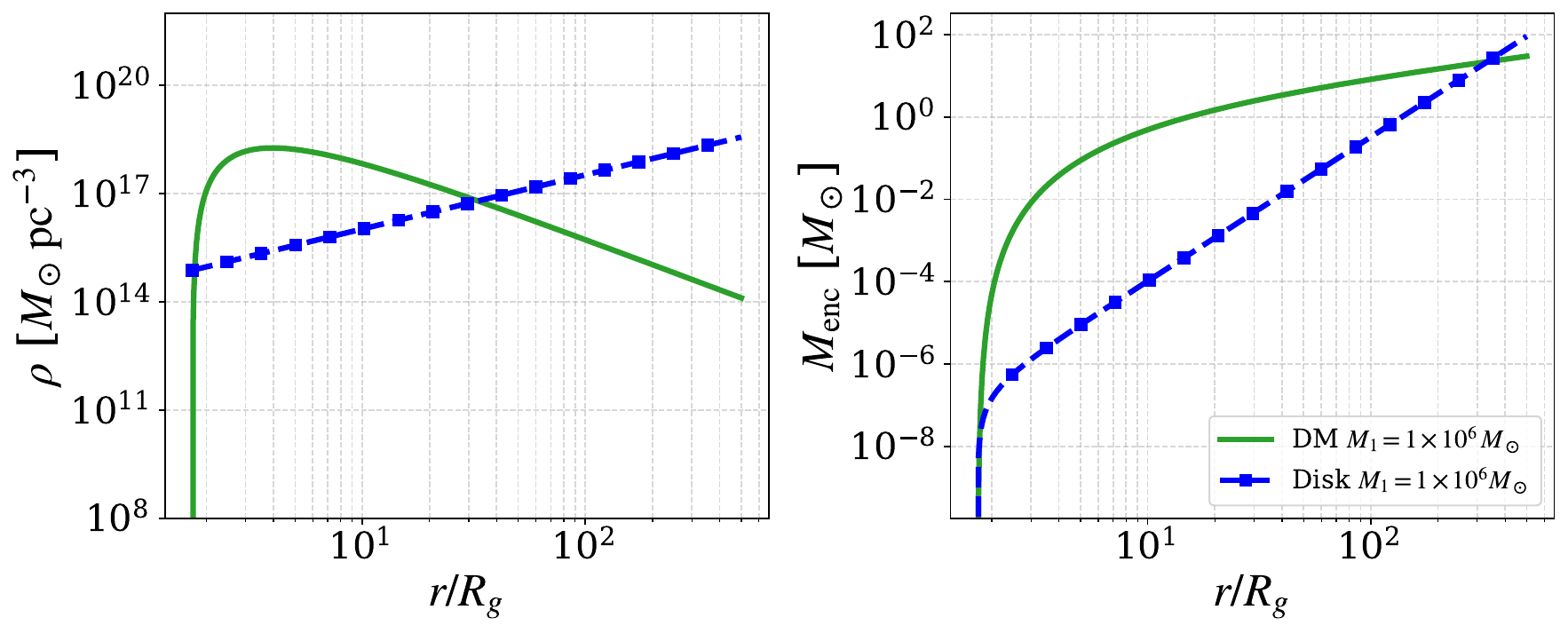}
    \caption[Density and enclosed mass profiles for DM spike and accretion disk]{
        Density (left) and enclosed mass $M_{\rm enc}(r)$ (right) as functions of radius for the DM spike (solid green) and the accretion disk (dashed blue) at the fiducial primary mass $M_1=10^6\,M_\odot$, spin $a=0.9$, and $\gamma=1$. The enclosed-mass curves cross at $r\simeq 3.3\times10^{2}\,R_g$, beyond which the disk enclosed mass surpasses that of the DM spike, and the enclosed DM mass within the orbital region ($p_0=10$) exceeds the disk mass by a factor of $\sim 4.8\times10^{3}$. Radii are in units of $R_g = GM_1/c^2$.
    }
    \label{fig:dm_vs_disk}
\end{figure}

FIG.~\ref{fig:dm_vs_disk} compares the DM spike and accretion disk profiles at the fiducial mass $M_1=10^6\,M_\odot$. The enclosed-mass comparison is directly relevant to the self-gravity channels, which are subdominant; the dominant effects---DM dynamical friction and disk torques---are instead governed by local quantities: the DM density $\rho_{\rm DM}(r)$ and the disk surface density $\Sigma(r)$, aspect ratio $h(r)$, and Mach number $e/h$. These two effects are therefore sensitive to different radial scales: DM dynamical friction dominates at small separations where the spike density is highest, while disk torques become competitive at larger semi-major axes, where the longer inspiral duration allows the gas torque to accumulate despite the lower orbital density. The single disk profile reflects the treatment of $\Sigma_0$ and $h_0$ as free, mass-independent quantities; unlike the DM spike, whose density is tied to $M_1$ through the $M_1$--$\sigma_\star$ relation, the $\alpha$-disk normalization depends on the accretion parameters $\alpha$ and $f_{\rm Edd}$, whose uncertainties span orders of magnitude and dominate any residual mass trend.

\subsection{Gravitational-wave dephasing}

The 5PN AAK flux underlying the vacuum inspiral is an asymptotic series whose higher-order post-Newtonian terms overtake the leading order near the ISCO, where the flux develops a spurious zero above the physical separatrix. For numerical reliability, the dephasing quantities quoted below are computed from trajectories truncated at the point where the last (5PN) term reaches $10\%$ of the leading (0PN) term. For the fiducial system this truncates the inspiral at $p_{\rm cut}\simeq4.2$, corresponding to a truncation time $t_{\rm cut}\simeq3.6$ yr of the four-year observation. Varying this threshold to $20\%$ or $30\%$ changes the quoted dephasing by at most a factor of $\sim2$, without altering the qualitative conclusions. The mismatch and Fisher-matrix diagnostics below are instead computed on the full trajectories, since their qualitative conclusions are insensitive to this truncation.

\begin{figure}[t]
    \centering
    \includegraphics[width=\textwidth]{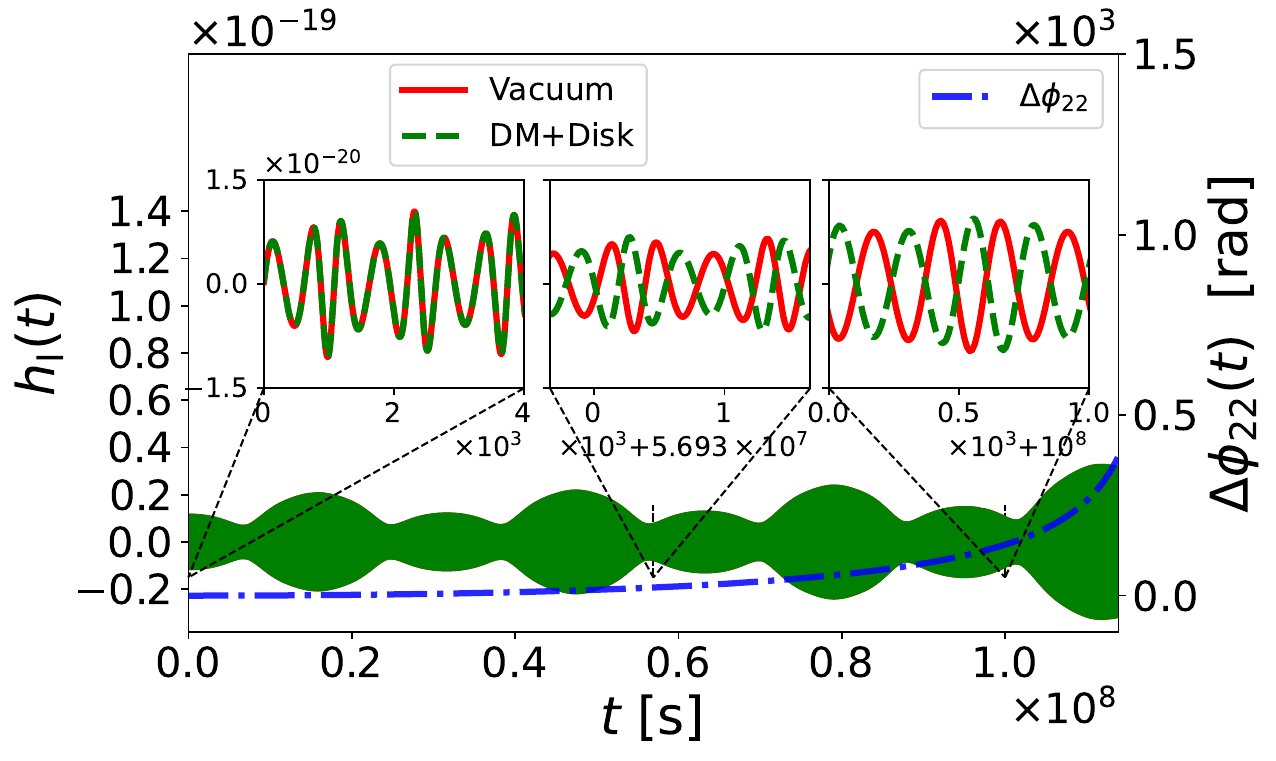}
    \caption[Waveform comparison: vacuum versus all environmental effects]{
        Detector-frame strain $h_{\rm I}(t)$ for the fiducial EMRI system ($M_1=10^6\,M_\odot$, $m_2=10\,M_\odot$, $a=0.9$, $p_0=10$, $e_0=0.2$, $\gamma=1$) over a four-year observation. The red solid curve shows the vacuum (5PN Kerr) waveform; the green dashed curve (labeled DM$+$Disk) includes both effects simultaneously. The blue dash-dotted curve (right axis) shows the cumulative $(2,2)$-harmonic dephasing $\Delta\phi_{22}(t) = 2[\Phi_\phi^{\rm env}(t) - \Phi_\phi^{\rm vac}(t)]$, which reaches $\sim 4\times10^2\,\mathrm{rad}$ at the end of the observation. The three inset panels zoom into the early, middle, and late stages of the inspiral, illustrating the gradual accumulation of the phase difference.
    }
    \label{fig:waveform_comparison}
\end{figure}

To build intuition for how environmental effects manifest in the waveform, we first compare the time-domain detector strain for the fiducial system with all environmental modules active against the vacuum case. FIG.~\ref{fig:waveform_comparison} shows that the environmental perturbations accumulate as a phase drift. The drift accelerates at late times as the secondary samples regions of higher environmental density and stronger gravitational field. This dephasing $\Delta\phi_{22}(t)$ is the primary diagnostic we quantify across the parameter space in the remainder of this subsection.

\begin{figure}[t]
    \centering
    \includegraphics[width=1\textwidth, height=0.85\textwidth]{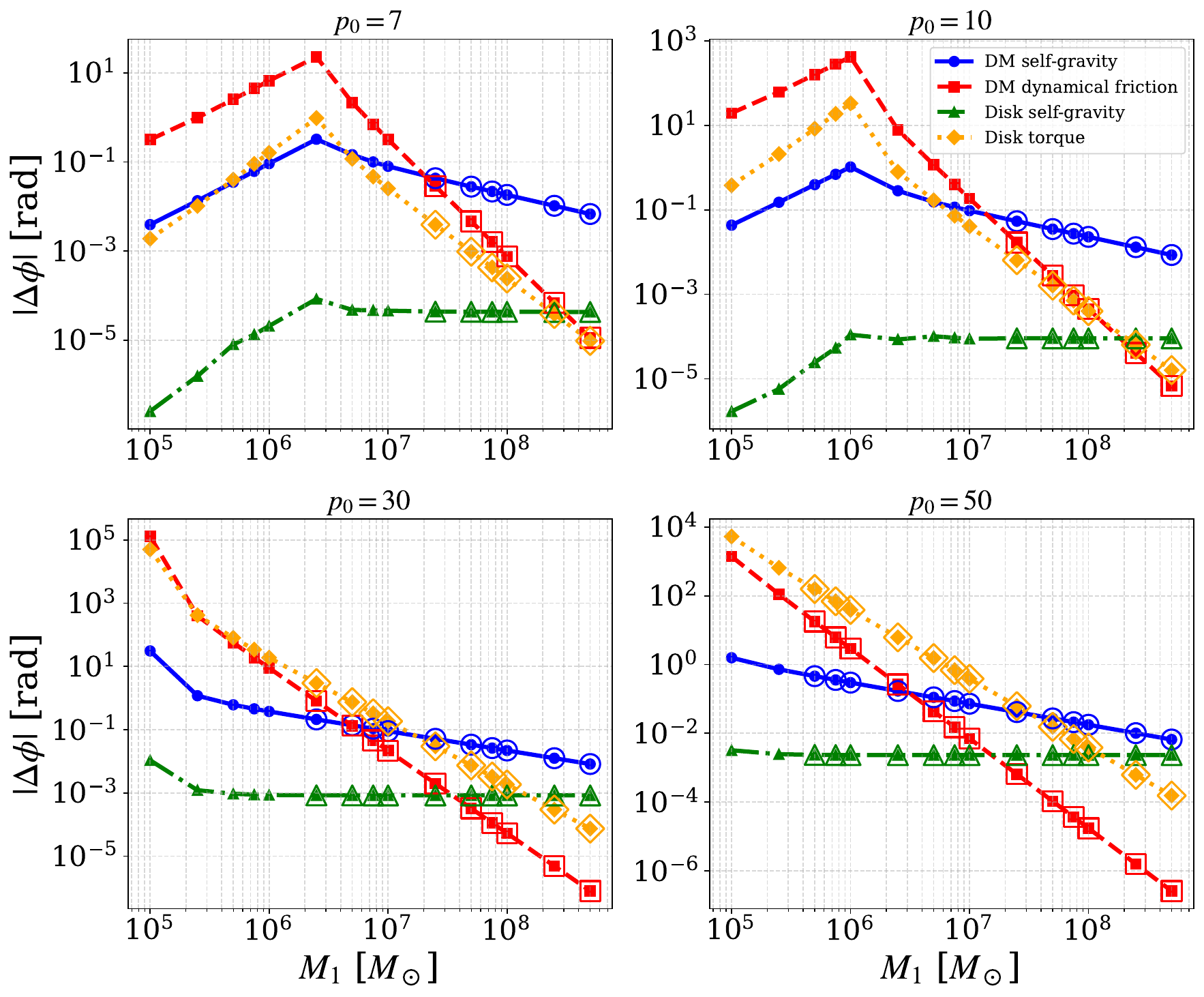}
    \caption[Dephasing vs primary mass for different $p_0$]{$(2,2)$-harmonic dephasing $|\Delta\phi_{22}|$ as a function of primary mass $M_1$ for $p_0=7,10,30,50$ (left to right, top to bottom), at $a=0.9$, $e_0=0.2$, and $\gamma=1$. The four curves in each panel correspond to DM self-gravity (solid blue), DM dynamical friction (dashed red), disk self-gravity (dash-dotted green), and disk torque (dotted orange). Hollow markers indicate data points with SNR $<10$.}
    \label{fig:dephasing_compare_p0}
\end{figure}

FIG.~\ref{fig:dephasing_compare_p0} maps the dephasing as a function of $M_1$ at four initial separations. The DM dynamical-friction dephasing is non-monotonic in mass, peaking at intermediate $M_1$ and falling off toward both lighter and heavier primaries. The peak reflects a competition: lighter systems inspiral for longer geometrically and accumulate more orbital cycles, yet their absolute observation window shrinks, while the $M_1$--$\sigma_\star$ relation makes the DM spike denser around lighter black holes (FIG.~\ref{fig:dm_vs_disk}). At high masses the brief inspiral suppresses the dissipative channels, and the conservative self-gravity effects---whose frequency shifts accumulate regardless of the orbital cycle count---dominate at a low dephasing level. The DM and disk contributions also respond oppositely to $p_0$: the DM dephasing weakens with increasing separation as the spike density declines, whereas the disk torque strengthens and dominates at large separations, except at the lightest masses, where the smallest initial separation starts nearest the separatrix and reverses the trend.

\begin{figure}[t]
    \centering
    \includegraphics[width=1\textwidth, height=0.35\textwidth]{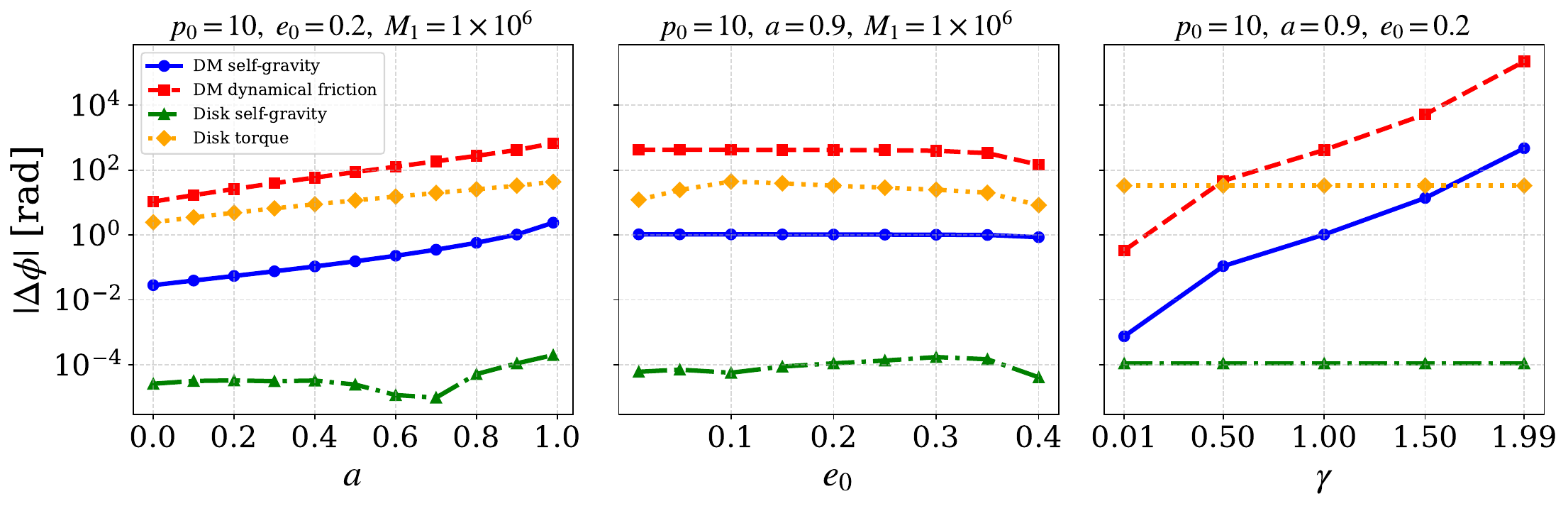}
    \caption[Dephasing vs spin, eccentricity, and DM slope]{$(2,2)$-harmonic dephasing $|\Delta\phi_{22}|$ as a function of primary spin $a$ (left), initial eccentricity $e_0$ (middle), and DM spike slope $\gamma$ (right), at $M_1=10^6\,M_\odot$ and $p_0=10$, with the non-scanned parameters held at their fiducial values. Line styles: DM self-gravity (solid blue), DM dynamical friction (dashed red), disk self-gravity (dash-dotted green), disk torque (dotted orange).}
    \label{fig:dephasing_compare_a_e0_gamma}
\end{figure}

FIG.~\ref{fig:dephasing_compare_a_e0_gamma} isolates the dephasing dependence on the Kerr spin, orbital eccentricity, and the spike-slope. The DM dynamical-friction and disk-torque dephasing both grow substantially with spin, driven by the longer inspiral of higher-spin systems; the disk grows somewhat less steeply, since the Newtonian torque is insensitive to the Kerr geometry and inherits only the inspiral-duration dependence. The DM self-gravity channel tracks the same spin trend at a much lower amplitude and remains nearly constant with eccentricity. Across eccentricity, the DM dynamical-friction dephasing varies only mildly---the enhanced pericentre velocity being offset by the larger semi-major axis---and its apparent decline at the largest eccentricity is a truncation artifact. The disk-torque dephasing, by contrast, changes sign across the transonic transition, peaks near $e_0\sim h_0$, and falls off at larger eccentricity as the supersonic $(h/e)^3$ suppression sets in. Overall, the subsonic-to-supersonic transition does not imprint a sharp feature in the dephasing magnitude that distinguishes the DM and disk channels.

The DM dephasing is, however, exquisitely sensitive to the spike slope $\gamma$: both the dynamical friction and self-gravity effects span five to six orders of magnitude across $\gamma\in[0.01,1.99]$. This extreme sensitivity reflects the adiabatic-contraction mapping: the modest variation of $\xi=(9-2\gamma)/(4-\gamma)$ is amplified by the nine-decade radial separation between the orbital region and the halo scale $r_0$, and the $\gamma$-dependence of the spike normalization $(\rho_{\rm sp},r_{\rm sp})$ (Eqs.~\eqref{eq:rhosp}--\eqref{eq:rsp}) contributes an additional three orders of magnitude. The disk dephasing is unaffected by $\gamma$ by construction.

The complementary mismatch diagnostics (Sec.~\ref{sec:analysis}) confirm that these dephasing levels translate into statistically distinguishable waveform differences. The DM dynamical-friction mismatch reaches $\mathcal{M}\sim 0.6$--$0.99$ across the parameter space, while the DM self-gravity mismatch attains $\mathcal{M}\sim 0.02$--$0.8$, well above the LISA distinguishability threshold of $\mathrm{SNR}^{-2}\approx 4\times 10^{-4}$ at the fiducial SNR of 50; both DM channels are therefore unambiguously detectable. The disk torque mismatch is comparably significant, confirming that the torque signal can be distinguished from vacuum even where its absolute dephasing is moderate. Disk self-gravity remains the only channel whose mismatch falls below the threshold across all configurations.

\begin{figure}[t]
    \centering
    \includegraphics[width=\textwidth]{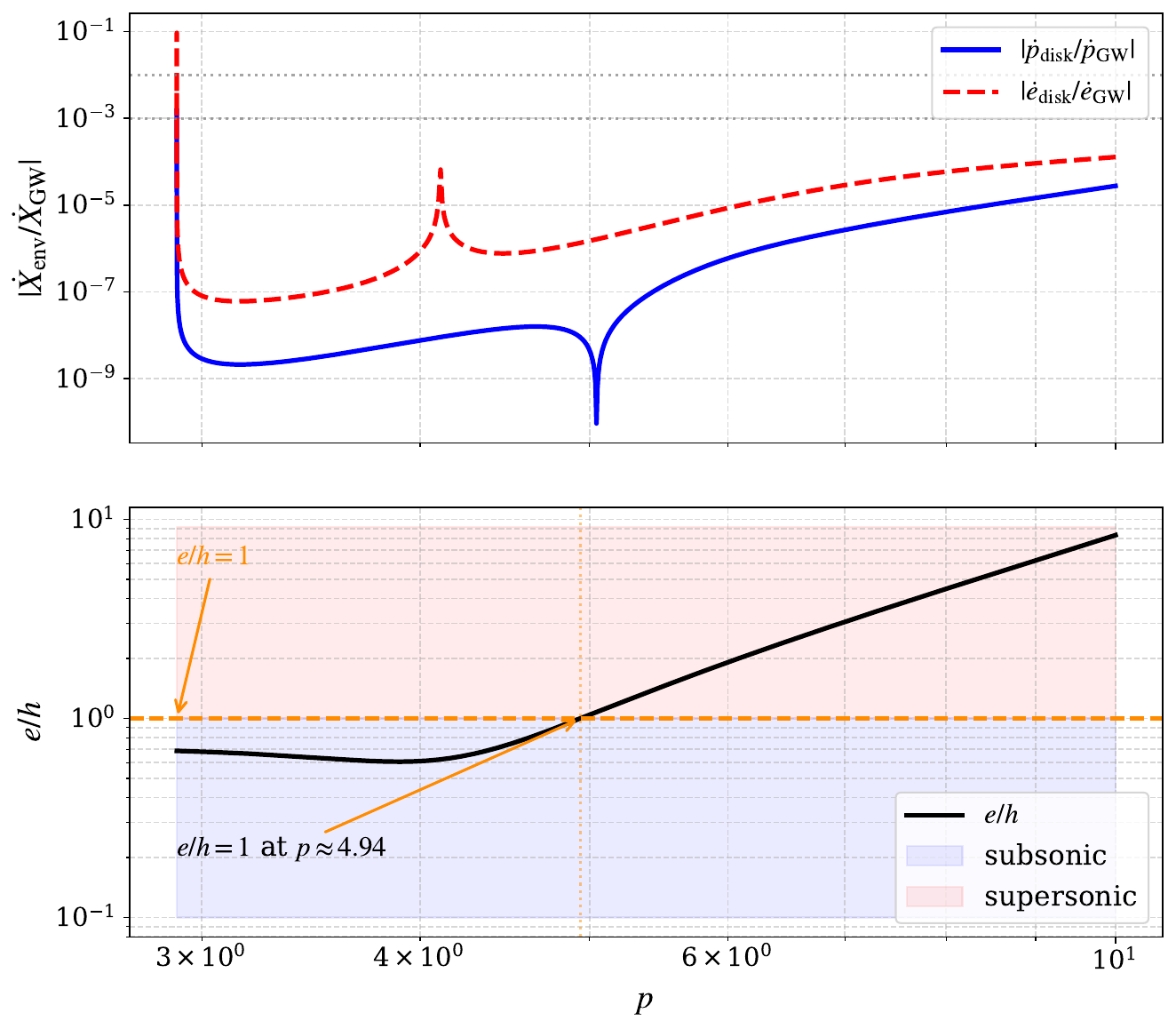}
    \caption[Adiabatic diagnostic and subsonic-to-supersonic transition]{
        Disk torque adiabatic diagnostic for a system with $M_1=10^6\,M_\odot$, $a=0.9$, $e_0=0.2$, and $p_0=10$. Upper panel: ratios $|\dot{p}_{\rm disk}/\dot{p}_{\rm GW}|$ (solid blue) and $|\dot{e}_{\rm disk}/\dot{e}_{\rm GW}|$ (dashed red), confirming that the disk torque is subdominant to the GW back-reaction, with median values $\sim 9\times10^{-6}$ and $\sim 7\times10^{-5}$, respectively. Lower panel: Mach number $e/h$ along the inspiral; the horizontal dash-dotted line marks $e/h=1$, with the subsonic ($e/h<1$) and supersonic ($e/h>1$) regimes shaded differently. The crossing at $p\simeq 4.9$ shows that the matching formulas of Eqs.~\eqref{eq:te_match}--\eqref{eq:ta_match} are sampled across the full transonic range.
    }
    \label{fig:adiabatic_disk}
\end{figure}

The validity of the orbit-averaging framework adopted throughout this work is assessed in FIG.~\ref{fig:adiabatic_disk}, which compares the disk torque with the GW back-reaction along the inspiral for the fiducial system. The torque ratios remain well below unity, confirming that the adiabatic treatment is self-consistent, and the crossing of $e/h=1$ confirms that the waveform samples both flow regimes; the implications of this transonic sampling for the measurability of $\Sigma_0$ and $h_0$ are addressed in the Fisher analysis below.

\subsection{Parameter estimation}

\begin{table}[t]
\centering
\caption{Fisher-matrix $1\sigma$ uncertainties and systematic biases, the latter in units of the $1\sigma$ statistical uncertainty ($\Delta/\sigma$), at a fixed detector-frame $\mathrm{SNR}=50$. For the disk and combined configurations the $1\sigma$ uncertainties for $\Sigma_0$ and $h_0$ are not reported, since their sub-matrix is near-singular (see text); the reported biases are derived from the well-conditioned vacuum Fisher matrix. $\|\Delta h_{\rm env}\|\equiv\sqrt{\langle h_{\rm env}-h_{\rm vac}|h_{\rm env}-h_{\rm vac}\rangle}$ is the noise-weighted norm (i.e., the SNR) of the injected environmental waveform difference. A dash (---) indicates the absence of a parameter. All condition numbers are evaluated under log-parameter reparameterization (Sec.~\ref{sec:analysis}).}
\label{tab:fisher}
\footnotesize
\begin{tabular}{lcccccccc}
\hline
\hline
 & \multicolumn{2}{c}{DM-only} & \multicolumn{2}{c}{Disk($e_0{=}0.2$, $a{=}0.9$)} & \multicolumn{2}{c}{Disk($e_0{=}0.03$, $a{=}0.04$)} & \multicolumn{2}{c}{Combined} \\
 & ($\sigma$) & ($\Delta/\sigma$) & ($\sigma$) & ($\Delta/\sigma$) & ($\sigma$) & ($\Delta/\sigma$) & ($\sigma$) & ($\Delta/\sigma$) \\
\hline
$M_1$          & $2.57{\times}10^{-1}$ & $+0.4$   & $2.63{\times}10^{-1}$ & $+0.5$   & $4.97{\times}10^{2}$  & $-0.00$ & $2.58{\times}10^{-1}$ & $+0.2$ \\
$m_2$          & $9.22{\times}10^{-6}$ & $-22.4$  & $4.65{\times}10^{-5}$ & $-0.5$   & $1.60{\times}10^{-3}$ & $-0.01$ & $2.20{\times}10^{-5}$ & $+8.2$ \\
$a$            & $2.79{\times}10^{-7}$ & $+2.1$   & $2.79{\times}10^{-7}$ & $-3.7$   & $6.65{\times}10^{-4}$ & $-1.2$  & $2.80{\times}10^{-7}$ & $-2.3$ \\
$p_0$          & $4.85{\times}10^{-6}$ & $+1.1$   & $5.54{\times}10^{-6}$ & $-11.4$  & $2.84{\times}10^{-3}$ & $-0.00$ & $5.36{\times}10^{-6}$ & $-5.0$ \\
$e_0$          & $1.23{\times}10^{-6}$ & $-6.4$   & $1.49{\times}10^{-6}$ & $+22.3$  & $1.95{\times}10^{-3}$ & $-0.03$ & $1.49{\times}10^{-6}$ & $+9.4$ \\
\hline
$\gamma$       & $3.34{\times}10^{-3}$ & ---      & --- & --- & --- & --- & $4.89{\times}10^{-3}$ & --- \\
\hline
$\|\Delta h_{\rm env}\|$ & \multicolumn{2}{c}{$57$} & \multicolumn{2}{c}{$77$} & \multicolumn{2}{c}{$1.8$} & \multicolumn{2}{c}{$52$} \\
$\kappa(\Gamma)$ & \multicolumn{2}{c}{$1.69{\times}10^{13}$} & \multicolumn{2}{c}{$4.25{\times}10^{18}$} & \multicolumn{2}{c}{$1.04{\times}10^{20}$} & \multicolumn{2}{c}{$6.75{\times}10^{17}$} \\
\hline
\end{tabular}
\end{table}

\begin{figure}[t]
    \centering
    \includegraphics[width=\textwidth]{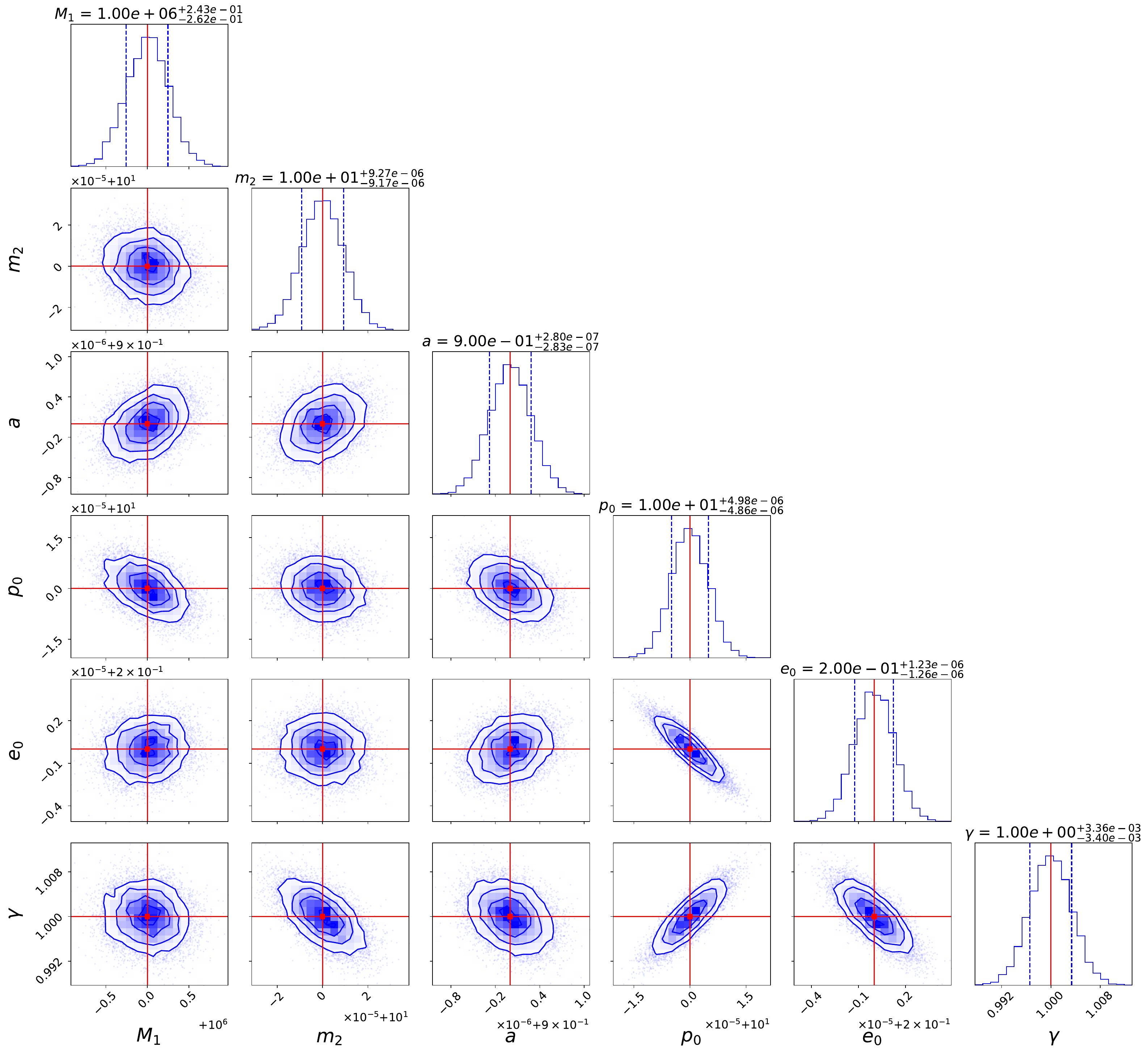}
    \caption[Fisher-matrix corner plot for the DM-only configuration]{
        Corner plot of the DM-only Fisher analysis for the fiducial system ($M_1=10^6\,M_\odot$, $m_2=10\,M_\odot$, $a=0.9$, $p_0=10$, $e_0=0.2$, $\gamma=1$, SNR$=50$). The six parameters are $M_1$, $m_2$, $a$, $p_0$, $e_0$, and $\gamma$. The diagonal panels show the marginalized one-dimensional Gaussian Fisher likelihoods; the off-diagonal panels show the two-dimensional $1\sigma$ and $2\sigma$ Gaussian Fisher contours. A moderate correlation between $a$ and $\gamma$ ($\mathcal C_{a,\gamma}\simeq -0.29$) is visible.
    }
    \label{fig:fisher_dm_corner}
\end{figure}

\begin{figure}[t]
    \centering
    \includegraphics[width=\textwidth]{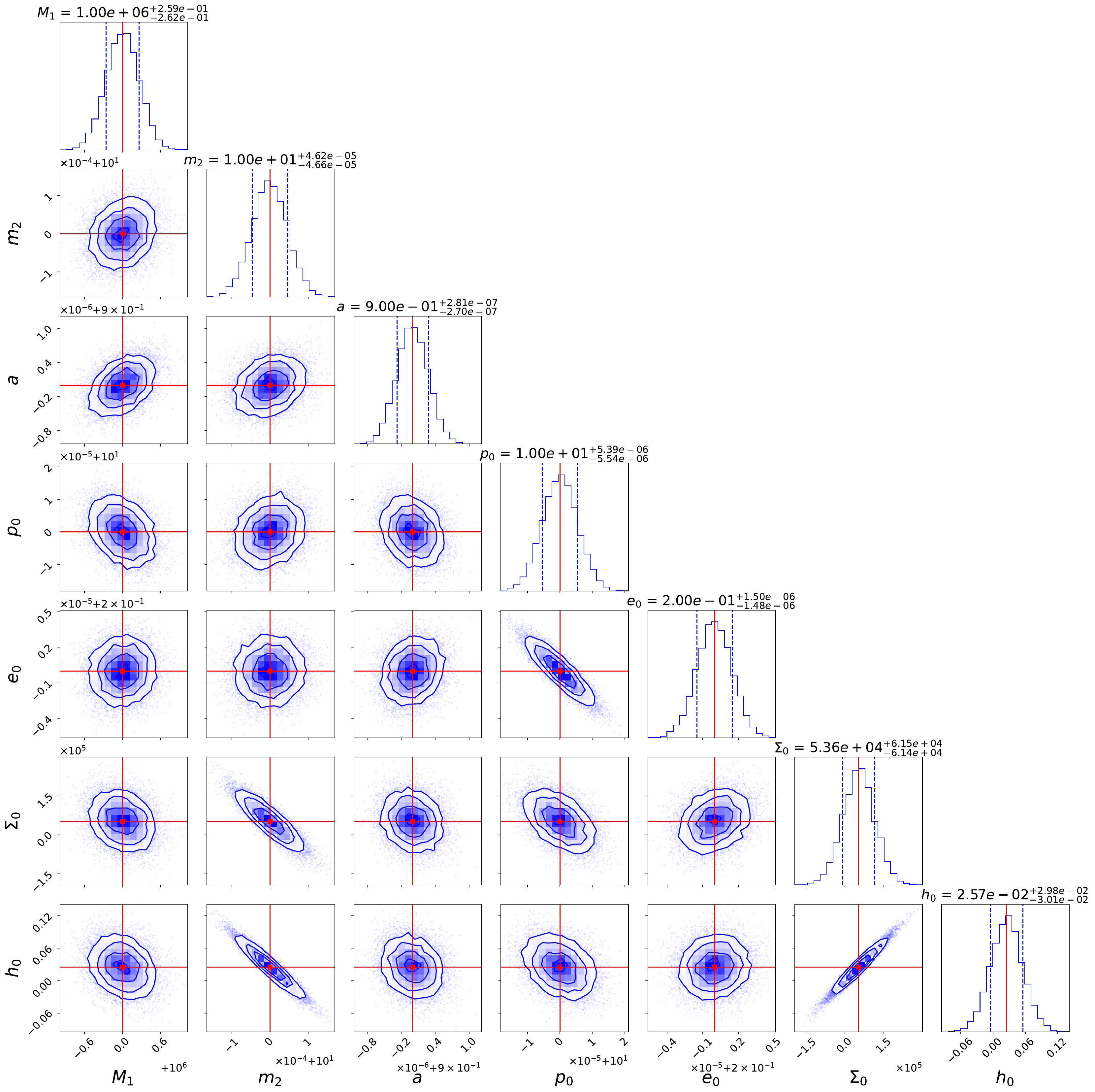}
    \caption[Fisher-matrix corner plot for the disk-only configuration]{
        Corner plot of the disk-only Fisher analysis for $M_1=10^6\,M_\odot$, $m_2=10\,M_\odot$, $a=0.9$, $p_0=10$, $e_0=0.2$, $\Sigma_0=5.25\times10^{4}\,\mathrm{g\,cm^{-2}}$, $h_0=0.025$, SNR$=50$. The diagonal and off-diagonal panels show the marginalized Gaussian Fisher likelihoods; the near-perfect correlation between $\Sigma_0$ and $h_0$ reflects the degeneracy discussed in the text.
    }
    \label{fig:fisher_disk_corner}
\end{figure}

\begin{figure}[t]
    \centering
    \includegraphics[width=\textwidth]{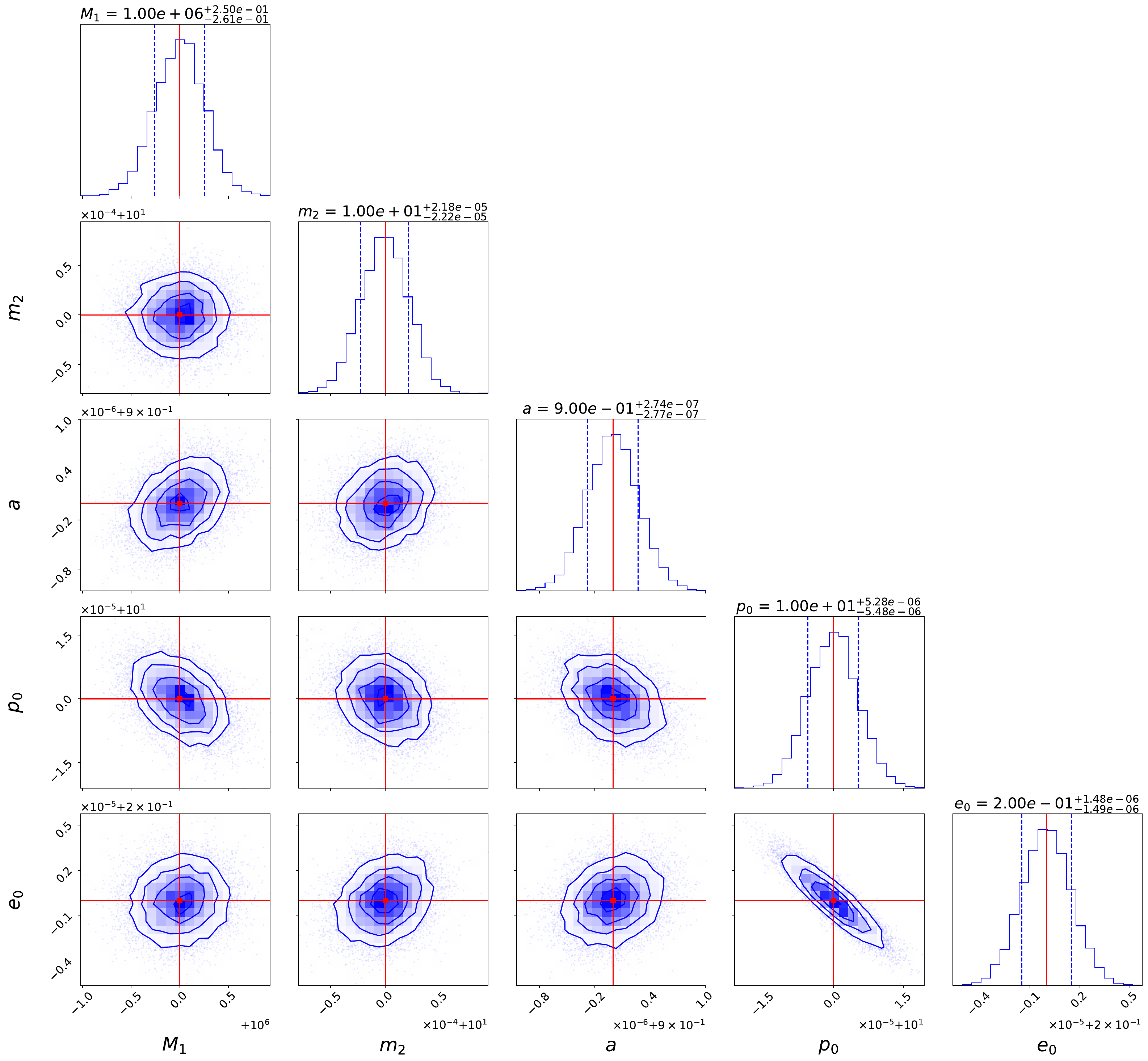}
    \caption[Fisher-matrix corner plot for the combined configuration: intrinsic parameters]{
        Corner plot of the combined DM$+$disk Fisher analysis for the fiducial system ($M_1=10^6\,M_\odot$, $m_2=10\,M_\odot$, $a=0.9$, $p_0=10$, $e_0=0.2$, $\Sigma_0=5.25\times10^{4}\,\mathrm{g\,cm^{-2}}$, $h_0=0.025$, $\gamma=1$, SNR$=50$). The five intrinsic EMRI parameters $M_1$, $m_2$, $a$, $p_0$, and $e_0$ are shown. The diagonal panels show the marginalized one-dimensional Gaussian Fisher likelihoods; the off-diagonal panels show the two-dimensional $1\sigma$ and $2\sigma$ Gaussian Fisher contours.
    }
    \label{fig:fisher_combined_intrinsic}
\end{figure}

\begin{figure}[t]
    \centering
    \includegraphics[width=\textwidth]{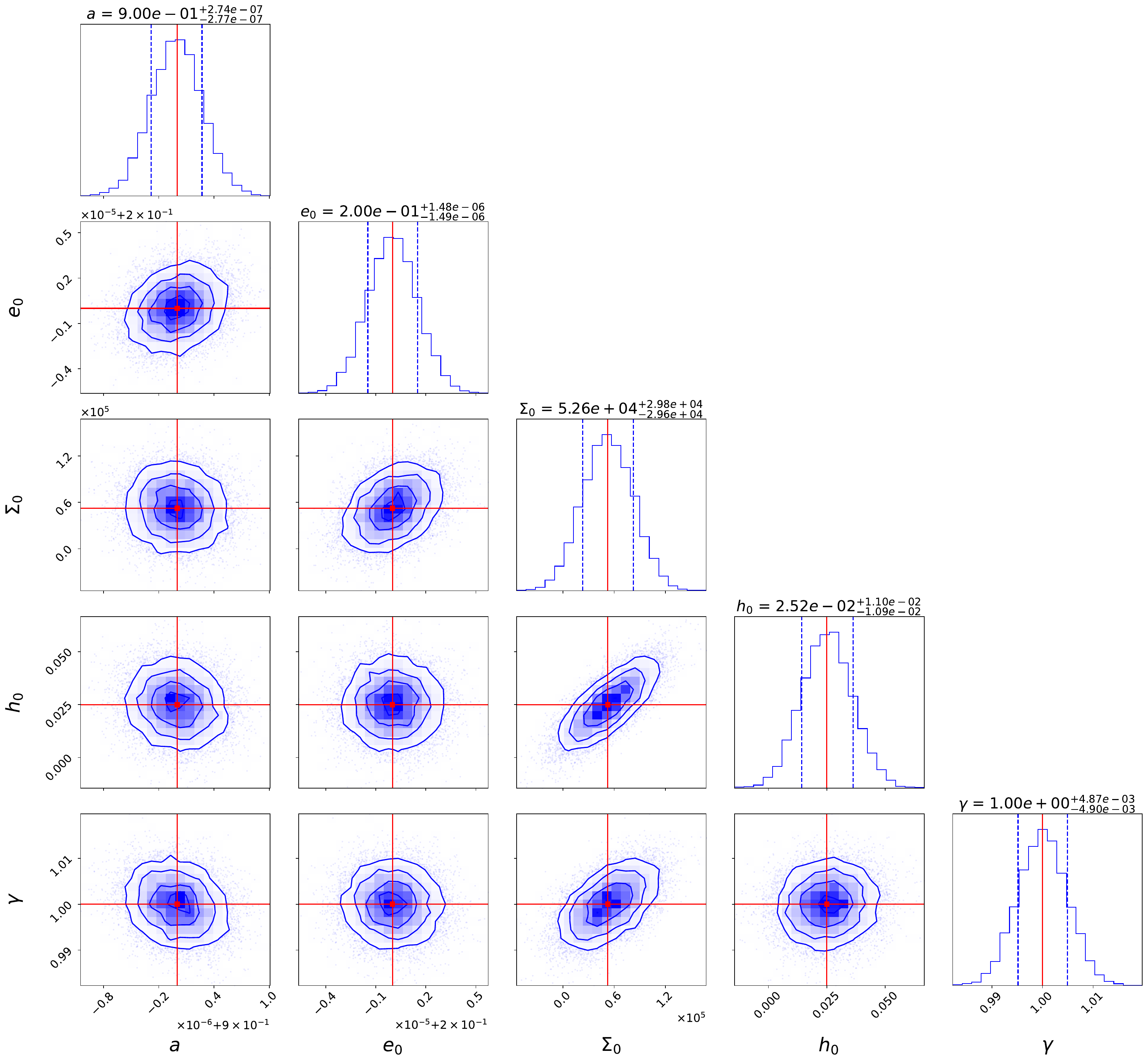}
    \caption[Fisher-matrix corner plot for the combined configuration: environmental parameters]{
        Same combined Fisher analysis as in FIG.~\ref{fig:fisher_combined_intrinsic}, showing the environmental and associated parameters $a$, $e_0$, $\Sigma_0$, $h_0$, and $\gamma$. The cross-sector correlation between $\gamma$ and $\Sigma_0$ is visible.
    }
    \label{fig:fisher_combined_env}
\end{figure}

\begin{figure}[t]
    \centering
    \includegraphics[width=\textwidth]{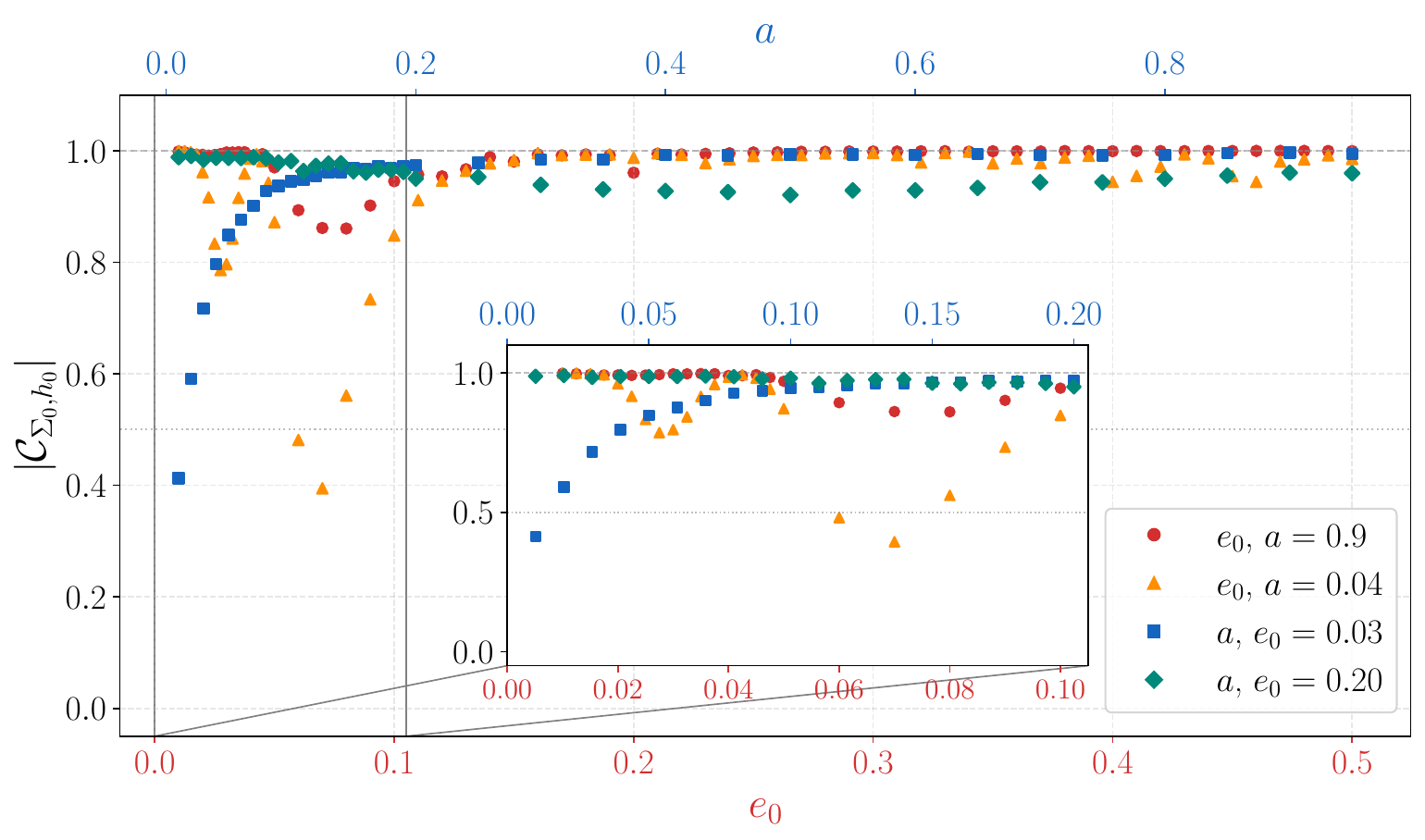}
    \caption[$\Sigma_0$--$h_0$ degeneracy scans across eccentricity and spin]{
        Correlation coefficient $|\mathcal C_{\Sigma_0,h_0}|$ from the disk-only Fisher analysis at SNR$=50$, evaluated at the stronger disk fiducial $\Sigma_0=5.25\times10^{5}\,\mathrm{g\,cm^{-2}}$, $m_2=50\,M_\odot$, $p_0=16.83$, with $h_0=0.025$ and $\Sigma_p=-1.5$ held fixed. Bottom axis: scan over initial eccentricity $e_0$ at $a=0.9$ (red) and $a=0.04$ (orange); top axis: scan over Kerr spin $a$ at $e_0=0.03$ (blue) and $e_0=0.2$ (green). The degeneracy remains strong ($|\mathcal C_{\Sigma_0,h_0}|\gtrsim 0.9$) throughout, except in the transonic region at low spins, where it weakens only mildly, to $|\mathcal C_{\Sigma_0,h_0}|\simeq 0.4$---exhibiting two shallow dips, highlighted in the zoomed-in insets---without being lifted.
    }
    \label{fig:degeneracy_scan}
\end{figure}

We perform a Fisher-matrix analysis at a fixed detector-frame SNR of 50 (Sec.~\ref{sec:analysis}) for three configurations reflecting the modular structure of our implementation. The DM-only configuration has parameters $\boldsymbol{\theta}_{\rm DM}=\{M_1, m_2, a, p_0, e_0, \gamma\}$ with the vacuum template obtained by switching off the DM module; the disk-only configuration has $\boldsymbol{\theta}_{\rm disk}=\{M_1, m_2, a, p_0, e_0, \Sigma_0, h_0\}$, supplemented by scans over $e_0$ and $a$ to disentangle the roles of the Mach number and the Kerr geometry; and the combined configuration has $\boldsymbol{\theta}_{\rm comb}=\{M_1, m_2, a, p_0, e_0, \Sigma_0, h_0, \gamma\}$, probing cross-sector degeneracies. Biases on the vacuum parameters are estimated via Eq.~\eqref{eq:bias}.

TABLE.~\ref{tab:fisher} summarizes the Fisher-matrix results, with corner figures shown in FIGS.~\ref{fig:fisher_dm_corner}, \ref{fig:fisher_disk_corner}, \ref{fig:fisher_combined_intrinsic} and \ref{fig:fisher_combined_env}. In the DM-only configuration, the EMRI parameters are recovered with sub-percent precision and $\gamma$ is measurable to $\sim 0.33\%$; the dominant systematic bias appears in the secondary mass ($\Delta m_2\simeq -22\sigma$), reflecting the dissipative nature of the DM dynamical friction, which modifies the inspiral rate $\dot p,\dot e$ and is thus most degenerate with the mass-ratio scaling of the vacuum inspiral. A moderate correlation between $a$ and $\gamma$ ($\mathcal C_{a,\gamma}\simeq -0.29$) nonetheless persists, since the spin enters only geometrically through the spike inner edge $r_{\rm mb}$, partially overlapping with the slope dependence of $\gamma$ in shaping the density profile sampled by the orbit (FIG.~\ref{fig:fisher_dm_corner}).

The disk-only Fisher matrix is near-singular because of the degeneracy between $\Sigma_0$ and $h_0$ (FIG.~\ref{fig:fisher_disk_corner}), so the $1\sigma$ uncertainties of these two parameters are not reported, while the intrinsic EMRI parameters remain well determined. This degeneracy is physical rather than numerical (FIG.~\ref{fig:degeneracy_scan}) and traces back to the $e/h$ structure of the matched torque: in the supersonic limit both torque components scale as $\Sigma_0/h_0$, so the torques provide only one independent degree of freedom, while the subsonic branch ($1/t_e\propto\Sigma_0/h_0^4$ and $1/t_a\propto\Sigma_0/h_0^2$) carries the information separating $\Sigma_0$ from $h_0$. Since this separating information accumulates only while the orbit samples the transonic region, the degeneracy is weakened only mildly---to $\mathcal C_{\Sigma_0,h_0}\simeq0.4$---at near-Schwarzschild spins, where the crossing of $e/h\sim1$ occurs early in the slowly-evolving inspiral, and remains unbroken across the explored parameter space.

The combined DM$+$disk configuration introduces a moderate cross-sector degeneracy between $\gamma$ and $\Sigma_0$ (FIGS.~\ref{fig:fisher_combined_intrinsic} and \ref{fig:fisher_combined_env}), degrading the DM slope precision only mildly and shifting the vacuum-parameter biases relative to the DM-only case---the secondary-mass bias weakening while the eccentricity bias strengthens---reflecting the partially opposing distortions that the disk torque and the DM dynamical friction imprint on the recovered waveform; the near-singularity of the Fisher matrix nonetheless precludes reliable joint estimation of the individual environmental parameters. This degeneracy arises because $\gamma$ and $\Sigma_0$ are both amplitude-like parameters---controlling the DM density and the disk surface density---so that their dephasings accumulate secularly and overlap.

\section{Conclusions}\label{sec:conclusions}

We have presented a combined analysis of DM spike and accretion-disk environmental effects on eccentric, equatorial Kerr EMRIs, integrating both sectors as modular extensions to the 5PN AAK waveform model. For a $M_1=10^6\,M_\odot$ system, the DM dynamical-friction dephasing peaks at $M_1\sim 10^6\,M_\odot$, reaching $\sim 10^2$--$10^3\,\mathrm{rad}$, with the corresponding mismatch well above the LISA distinguishability threshold, and the DM spike slope $\gamma$ is measurable to sub-percent precision with a moderate $\gamma$--$a$ degeneracy; omitting DM effects incurs significant vacuum-template biases on the intrinsic parameters. DM self-gravity contributes non-negligibly across the full mass range, and the conservative self-gravity channels---for both the DM spike and the accretion disk---dominate over their dissipative counterparts at $M_1\gtrsim 10^8\,M_\odot$, because the frequency shifts they induce accumulate independently of the inspiral duration. The DM dephasing is exquisitely sensitive to the spike slope $\gamma$, with both the dynamical friction and self-gravity channels spanning five to six orders of magnitude, while the disk dephasing is insensitive to $\gamma$. The DM self-gravity mismatch also exceeds the detection threshold, reinforcing that both DM channels are unambiguously detectable.

In the accretion-disk sector, the surface density $\Sigma_0$ and aspect ratio $h_0$ remain degenerate across the explored parameter space, with $\mathcal C_{\Sigma_0,h_0}\simeq 1$ throughout most of it and only a mild weakening, to $\mathcal C_{\Sigma_0,h_0}\simeq 0.4$, at near-Schwarzschild spins; this degeneracy is physical, not numerical. The combined DM$+$disk analysis further reveals a cross-sector degeneracy between $\gamma$ and $\Sigma_0$. Extending this framework to general inclined orbits may introduce an additional dependence through the inclination relative to the disk mid-plane, possibly reshaping the degeneracy structure. Relativistic torque models and electromagnetic counterpart observations for Kerr EMRIs may provide a new estimation framework that breaks this degeneracy and yields complete estimates of the accretion-disk parameters.

\begin{acknowledgments}
This work is partly supported by the National Key Research and Development Program of China (Grant No.~2021YFC2201901), the National Natural Science Foundation of China (NSFC) under Grants Nos. 12547104, and the Fundamental Research Funds for the Central Universities.
\end{acknowledgments}

\bibliographystyle{unsrtnat}
\bibliography{references}

\end{document}